\documentstyle[twocolumn,prl,aps]{revtex}
\large
\begin{document}
\draft
\twocolumn[
\hsize\textwidth\columnwidth\hsize\csname @twocolumnfalse\endcsname

\title
      {
        Acoustic Nature of the Boson Peak in Vitreous Silica
      }
\author{
        C.~Masciovecchio$^{1}$, 
        V.~Mazzacurati$^{2}$, 
        G.~Monaco$^{2}$, 
        G.~Ruocco$^{2}$, 
        T.~Scopigno$^{2}$, 
        F.~Sette$^{1}$, 
        P.~Benassi$^{2}$, 
        A.~Cunsolo$^{1}$, 
        A.~Fontana$^{3}$, 
        M.~Krisch$^{1}$, 
        A.~Mermet$^{1}$, 
        M.~Montagna$^{3}$, 
        F.~Rossi$^{3}$, 
        M.~Sampoli$^{4}$, 
        G.~Signorelli$^{2}$, 
        R.~Verbeni$^{1}$
                }
\address{
         $^1$
         European Synchrotron Radiation Facility, BP 220, 
         F-38043, Grenoble, Cedex, France. \\
         $^2$
         Dipartimento di Fisica and INFM, Universit\`a di L'Aquila, 
         I-67100, L'Aquila, Italy. \\
         $^3$
         Dipartimento di Fisica and INFM, Universit\'a di Trento, 
         I-38100, Trento, Italy.\\
         $^4$
         Dipartimento di Energetica and INFM, Universit\'a di Firenze, 
         I-50139, Firenze, Italy.\\
        }

\date{\today }
\maketitle
\begin{abstract}
New temperature dependent inelastic x-ray (IXS) and Raman (RS) scattering
data are compared to each other and with existing inelastic neutron
scattering data in vitreous silica ($v-SiO_2$), in the 300 $\div$ 1775 K
region. The IXS data show collective propagating excitations up to $Q=3.5$ $%
nm^{-1}$. The temperature behaviour of the excitations at $Q=1.6$ $nm^{-1}$
matches that of the boson peak found in INS and RS. This supports the
acoustic origin of the excess of vibrational states giving rise to the boson
peak in this glass.
\end{abstract}


]

A peculiar characteristic of disordered solids, not present in their
crystalline counterpart (Phillips, 1981), is an excess of low frequency
states with respect to the predictions of the Debye theory. Inelastic
neutron scattering (INS) studies on the density of states, $g(\omega )$,
reveal, in fact, a broad peak (the {\it {boson peak}} ) (BP) in the quantity 
$g(\omega )/\omega ^2$ (Buchenau {\it et al}. 1986, Buchenau {\it et al}.
1988, Sokolov {\it et al}. 1995) at energies 2$\div $10 meV (15$\div $80 cm$%
^{-1}$), where the Debye theory would predict a constant. This excess is
responsible for the presence, in any glass studied so far, of a bump in the
temperature dependence of $C_p/T^3$ (Sokolov {\it et al.} 1993, Brodin {\it %
et al. }1994).

The broad boson peak feature is also observed in Raman scattering (RS)
spectra. This can be understood considering that the first order Raman
scattering intensity, $I(\omega )$, in the harmonic approximation, is
connected to $g(\omega )$ (Galeneer and Sen 1978): $I(\omega )=C(\omega
)g(\omega )[n(\omega ,T)+1]/\omega .$ Here $n(\omega ,T)$ is the Bose
population factor and $C(\omega )$ is the photon-excitation coupling
function, assumed to have a smooth behaviour in the considered energy region
(Fontana, Rocca and M. P.~Fontana 1987, Fontana {\it et al}. 1997).

The nature of this excess of states is still subject to speculation and, at
present, there are two different prevailing hypotheses (See Philosophical
Magazine, special issue: {\it V international workshop on disordered systems}%
, Andalo, 1994). In the first one, the excess of states is explained by the
localization of the high energy vibrational modes induced by the static
disorder in the glass. In the second case, collective propagating modes are
thought to persist at high frequency, and the BP reflects their density of
states. In the first view the BP energy corresponds to the energy of the
high $Q$ excitations, while in the second one the excitations are expected
to propagate at energies above the BP energy.

Information on the character of the excitations contributing to the boson
peak can be obtained from measurements of the dynamic structure factor $%
S(Q,\omega )$. Using inelastic x-ray scattering (IXS) (Sette {\it et al}.
1995), it has been possible to measure $S(Q,\omega )$ of different glasses
in the momentum-energy region of the BP, and to support the hypothesis of
the acoustic-like character of the collective modes and therefore of the
peak itself (Masciovecchio {\it et al}. 1996a, Masciovecchio {\it et al}.
subm.). In the specific case of $v$-silica, however, the IXS data have been
alternatively interpreted either with the phonon-localization model (Foret 
{\it et al}. 1997), or as propagating excitations (Benassi {\it et al }1997,
Masciovecchio {\it et al}. 1997).

A better insight on the relation between the excitations seen in $S(Q,\omega
)$ and the BP could be gained by comparing the temperature dependence of
their energies. The basic idea is to utilise the observation that in
vitreous silica, in contrast to the majority of glasses, the low frequency
sound velocity, measured by Brillouin Light Scattering (BLS), (Vacher and
Pelous 1976) and the boson peak energy (Wischnewski {\it et al. }1998) shift
towards higher values with increasing temperatures, and, more importantly,
the rates of change reported in the two cases are different. It is not the
aim of this work to account for these peculiarities, likely to be due to the
specific shape of the interatomic potential in silica. Rather, one can
utilize the different temperature dependence of excitations with different $Q
$, as seen by IXS, INS, RS and BLS, to establish a link between specific
vibrational states and the features of the density of states. In particular,
one can assess whether the excitations observed by IXS at energies
comparable to the BP energy have either a temperature behaviour similar to
that of the BP, or to that of the low $Q$ excitations measured by BLS, or
even another dependence.

In this Letter we report temperature dependent IXS and RS data of vitreous
silica. In the IXS spectra, propagating collective excitations are found at $%
T\approx $ 1400 K, thus confirming that the high frequency longitudinal
dynamics has an acoustic-like nature in the whole temperature region below
melting, and up to a momentum transfer $Q$=3.5 nm$^{-1}$, corresponding to
energies more than twice the BP energy ($E_{_{BP}}\approx 4$ meV at room
temperature). More importantly, we find the same temperature shift for the
BP energy and for the energy, $\Omega (Q)$, of the modes at $Q^{*}=1.6$ nm$%
^{-1}$ ($\Omega (Q^{*})\approx E_{_{BP}}$). This supports the argument that
the acoustic modes, probed by the IXS, give a significant contribution to
the BP.

The IXS experiment was carried out at the very high energy resolution
inelastic x-ray scattering beamline (BL21/ID16) at the European Synchrotron
Radiation Facility. The total resolution function, measured using a
Plexiglas scatterer at the maximum of its static structure factor ($Q$=10~nm$%
^{-1}$), has a full width at half maximum ({\it fwhm}) of 1.5 $\pm $ 0.1
meV. The $Q$ values were selected between 0.75 and 4 nm$^{-1}$, and the $Q$
resolution was set to 0.2 nm$^{-1}$ {\it fwhm}. Energy scans were performed
by varying the relative temperature between the monochromator and the
analyser crystals by $\pm $0.45~K with a step of 0.0075 K. Each scan took
about 200 min, and each $Q$-value was obtained by averaging five scans (the
total integration time was 500~s per point). The data were normalized to the
intensity of the incident beam. Further details on the IXS beamline are
reported elsewhere (Masciovecchio {\it et al. }1996a, b, Verbeni {\it et al}%
. 1996). The Raman scattering measurements were performed using a standard
Raman laser system. Depolarized spectra were collected in the -300$\div $%
2000 cm$^{-1}$ frequency range with 1 cm$^{-1}$ frequency resolution. The $%
SiO_2$ suprasil sample, purchased from Goodfellow, was the same 2 $mm$
diameter rod used in (Benassi {\it et al }1997, Masciovecchio {\it et al}.
1997).

From the linear dispersion curve determined by a first set of IXS spectra
taken at $T$=1375 K, it was possible to establish that the excitations have
the same propagating nature as that previously observed in the 300$\div $%
1000 K region. In Fig.~1, we report as an example the IXS spectrum taken at $%
Q=1$ nm$^{-1}$, together with the fitting function. As in previous works,
the IXS spectra were fitted by the experimentally determined resolution
function convoluted with a $\delta $-function for the elastic peak and a
Damped Harmonic Oscillator (DHO) model (Fak and Dorner, 1992) for the
inelastic signal. The excitation energies are reported in the inset. The
data show an acoustic like behaviour, linearly extrapolating at small $Q$
with a slope $v=6800\pm 200$ m/s. This value, larger than the sound velocity
measured by Brillouin light scattering (Vacher {\it et al.} 1976) implies a
deformation of the acoustic branch not recognized before, which will provide
an interesting subject to further investigations. The dispersion relation
covers the whole BP energy region ($\approx 5\div 7$ meV), and reaches
energies as high as 15$\div $20 meV. These observations, as in previous
lower temperature measurements, show that, even at temperatures approaching $%
T_g$, it exists an acoustic-like propagating mode up to energies that lie
well above $E_{_{BP}}$.

A second set of IXS measurements was performed as a function of temperature
at the fixed value $Q^*=1.6$ nm$^{-1}$, which was chosen since $\Omega(Q^*)$
is in the range of $E_{_{BP}}$ at room temperature. The IXS data and the
corresponding fits are shown in Fig.~2. It is possible to observe directly
from the row data that the intensity and energy of the inelastic excitation
increase with $T$. This is emphasized in the inset of Fig.~2, where the $T$%
-dependence of the inelastic-to-elastic intensity ratio, $R(T)$, and the
excitation energy, $\Omega(Q^*,T)$, are reported.

To compare the $T-$dependence of $\Omega (Q^{*},T)$ with that of other
spectroscopic features, we define the scaling factor $a_{_{IXS}}(T)$ as $%
\Omega (Q^{*},T)/\Omega (Q^{*},T=0K)$. Here $\Omega (Q^{*},T=0K)$ is the $T=0
$ extrapolation of the measured $\Omega (Q^{*},T)$. The values of $%
a_{_{IXS}}(T)$ are reported in Fig.~3, together with the scaling factor for
the low $Q$ ($Q\approx 0.036$ nm$^{-1}$) excitation energy determined by
Brillouin light scattering (Scopigno 1997), $a(T)_{_{BLS}}$. The difference
in the $T$-dependence of the low- and high-$Q$ excitations indicates again
that the acoustic modes dispersion relation is deformed from a simple linear
law. This behaviour was hidden in the error bars in previous IXS and BLS
studies (Benassi {\it et al }1997, Masciovecchio {\it et al}. 1997). and is
evidenced here thanks to increased sensitivity and temperature range. It is
not the aim of the present work to discuss the origin of the observed
deformation of the dispersion relation \footnote{%
This effect could, for example, be ascribed to the presence of a positive
fourth order anharmonicity in the next-nearest neighbours interaction
potentials (Scopigno 1997) or to a relaxation process active in the glass.} .

This non trivial $T$-dependence can help us to assess the origin of the BP
by the study of the temperature dependence of its energy. In Fig.~3 we
report, in fact, the scaling factor $a_{_{INS}}(T)$ of the BP energy
maximum, as determined from existing INS measurements (Wischnewski {\it et al%
}. 1998). The striking similarity between $a_{_{INS}}(T)$ and $a_{_{IXS}}(T)$
provides a first strong indication that the states probed by IXS contribute
to the BP. This similarity is further confirmed by the temperature
dependence of the BP energy as measured by RS in the 300$\div $1100 K
region. In Fig.~4 we report $I(\omega )\omega /C(\omega )/[n(\omega ,T)+1]$,
which corresponds directly to $g(\omega )/\omega ^2$. We used the coupling
coefficient $C(\omega )$ derived from the ratio of the INS (Wischnewski {\it %
et al. }1998) and RS (this work) data measured in $v-SiO_2$ at room
temperature: this $C(\omega )$ is reported in the inset a) of Fig.~4, and it
was used to scale all the RS spectra at different temperatures \footnote{%
It has been recently shown that the coupling function is not dependent on
temperature in the energy region spanned by the BP, (A.~Fontana {\it et al.}
subm.).}. The energy of the maximum of the spectra of Fig.~4, derived by a
local quadratic fit, is reported in inset b) and the scaling factors $%
a_{_{RS}}(T)$ are reported in Fig.~3. The two data sets (INS and RS) are
equivalent within the error bar.

In presence of the deformation of the dispersion relation, it is striking to
observe the identical temperature behaviour for two spectroscopic features
that in principle may share only a similar energy: the BP maximum and the
acoustic excitations at $Q^*=$1.6 nm$^{-1}$. This demonstrates their common
origin. Consequently, the high frequency acoustic excitations, in spite of
their propagating nature, must have a density of states exceeding the Debye
prediction.

A model, that is consistent with a picture of linearly dispersing
propagating excitations with an excess of states, must take into account the
effect of the topological disorder on the eigenvectors of the vibrational
modes. It has been shown by molecular dynamics simulation that, in a glass
up to the BP energy region, the eigenvector of a given mode $j$ can be
thought as the sum of two very different components. {\it i)} A plane wave
like part, with an average momentum $Q_j$, related to the eigenvalue by the
linear dispersion $\omega _j\approx vQ_j$. The $Q$ distribution around $Q_j$
gets broader increasing the mode energy. {\it ii)} A random uncorrelated
component whose spatial Fourier transform extends from $Q_j$ up to $Q$
values much higher than the first sharp diffraction peak position with a
flat spectral distribution (Mazzacurati, Ruocco and Sampoli 1996). The first
component accounts for the peaks in the dynamic structure factor and for the
existence of a dispersion relation. The finite width of the inelastic peaks
in the $S(Q,\omega )$, as that in Fig. 1, is due to the finite projection of
eigenmodes of different energy at the considered $Q$. The presence of the
second component allows to satisfy the eigevectors orthogonality conditions
without the constrain on mode counting holding for plane waves, i.~e.
without imposing the $Q^2$ dependence of the modes density at a given
frequency. Specifically, with respect to the Debye behaviour, this can give
rise to an accumulation of states at low energy.

In conclusion, a study on the temperature dependence of the dynamics in
vitreous silica has allowed to identify equivalent temperature behaviours
for the maximum of the BP and for the high frequency continuation of the
sound branch at the same energy. This provides arguments in favour of the
acoustic-like origin for the excess in the density of states in vitreous
silica. Reminding that the BP is observed basically in any disordered
material, and that IXS has shown the presence of a propagating collective
dynamics in all the glasses and liquids studied so far, it is natural to
speculate on the generality of the conclusion reached here on the nature of
the BP.

We acknowledge A. P.~Sokolov and G.~Viliani for useful discussions.

\begin{center}
{\bf REFERENCES}
\end{center}

For a review see {\it Amorphous Solids : Low-Temperature properties}, edited
by W.A.~Phillips, (Springer, Berlin, 1981).

BENASSI P., KRISH M., MASCIOVECCHIO C., MONACO G., MAZZACURATI V., RUOCCO
G., SETTE F. and VERBENI R., 1997, {\it Phys. Rev. Lett.} {\bf 77}, 3835.

BRODIN A., FONTANA A., BORJESSON L., CARINI G. and TORELL L. M., 1994, {\it %
Phys. Rev. Lett.} {\bf 73}, 2067.

BUCHENAU U., PRAGER M., NUKER N., DIANOUX A. J., AHMAD N. and PHILLIPS W.
A., 1986, {\it Phys. Rev.} {\bf 34}, 5665.

BUCHENAU U., ZHOU H. M., NUKER N., GILROY K. S., and PHILLIPS W. A., 1988, 
{\it Phys. Rev. Lett.} {\bf 60}, 1318.

FAK B. and DORNER B., 1992, Institute Laue Langevin (Grenoble, France),
technical report No. 92FA008S.

FONTANA A., ROCCA F. and FONTANA M. P., 1987, {\it Phys. Rev. Letters} {\bf %
58}, 503.

FONTANA A., ROSSI F., CARINI G., D 'ANGELO G., TRIPODO G. and BARTOLOTTA A.,
1997, {\it Phys. Rev. Letters} {\bf 58}, 1078.

FONTANA A., DELL 'ANNA R., MONTAGNA M., ROSSI F., VILIANI G., RUOCCO G.,
SAMPOLI M., BUCHENAU U. and WISCHNEWSKI A., submitted to {\it Europhys.
letters}

FORET M., COURTENS E., VACHER R. and SUCK J. B., 1997, {\it Phys. Rev. Lett.}
{\bf 77}, 3831.

GALENEER F. L. and SEN P. N., 1978, {\it Phys. Rev. B}{\bf \ 17}, 1928.

MASCIOVECCHIO C., RUOCCO G., SETTE F., KRISH M., VERBENI R., BERGMAN U. and
SOLTWISCH M., 1996, {\it Phys. Rev. Lett.} {\bf 76}, 3356.

MASCIOVECCHIO C., BERGMAN U., KRISH M., RUOCCO G., SETTE F. and VERBENI R.,
1996, {\it Nucl. Inst. and Meth.} {\bf B-111}, 181 and {\bf B-117}, 339

MASCIOVECCHIO C., RUOCCO G., SETTE F., BENASSI P., CUNSOLO A, KRISH M.,
MAZZACURATI V., MERMET A., MONACO G. and VERBENI R., 1997, {\it Phys. Rev.} 
{\bf B55}, 8049.

MASCIOVECCHIO C., MONACO G., RUOCCO G., SETTE F., CUNSOLO A., KRISH M.,
MERMET A., SOLTWISCH M. and VERBENI R., submitted to {\it Phys. Rev. Lett}.

MAZZACURATI V., RUOCCO G. and SAMPOLI M., 1996, {\it Europhys. Lett.} {\bf 34%
}, 681.

SCOPIGNO T., 1997, Thesis, University of L'Aquila. Unpublished.

SETTE F., RUOCCO G., KRISH M., BERGMAN U., MASCIOVECCHIO C., MAZZACURATI V.,
SIGNORELLI G. and VERBENI R., 1995, {\it Phys. Rev. Lett.} {\bf 75}, 850.

SOKOLOV A. P., BUCHENAU U., STEFFEN W., FRICK B. and WISCHNEWSKI A., 1995, 
{\it Phys. Rev. B}{\bf \ 52}, R9815.

SOKOLOV A. P., KISLIUK A., QUITMANN. D. and DUVAL E., 1993, {\it Phys. Rev. B%
}{\bf \ 48}, 7692.

See for example {\it Philosophical Magazine, }1995{\it ,} {\it \ }{\bf B 71}%
, 471-811. Special issue: {\it V international workshop on disordered systems%
}, Andalo, 1994, A.~Fontana and G.~Viliani Editors.

VACHER R. and PELOUS J., 1976, {\it Phys. Rev.} {\bf B14}, 823.

VERBENI R., SETTE F., KRISH M., BERGMAN U., GORGES B., HALCOUSSIS C., MARTEL
K., MASCIOVECCHIO C., RIBOIS J.F., RUOCCO G. and SINN H., 1996, {\it J. of
Synchrotron Radiation} {\bf 3}, 62.

WISCHNEWSKI A., BUCHENAU U., DIANOUX A. J., KAMITAKARHA W. A. and ZARESTKY
J. L., 1998, {\it Phil. mag.} B77, 579.

\newpage\ 

{\footnotesize {\ }}

\begin{center}
{\footnotesize {\bf FIGURE CAPTIONS} }
\end{center}

\begin{description}
\item  {\footnotesize {FIG. 1 - Inelastic x-ray scattering spectrum of $%
v-SiO_2$ at T=1375 K and $Q=1$ $nm^{-1}$ ($\circ $). The full line is the
best fit to the data as discussed in the text. The dashed and dotted lines
represent the elastic and inelastic contributions to the fit. The inset
reports the energy of the excitations as derived from the fits. The dashed
line is the linear extrapolation to small $Q$, and its slope corresponds to
6800 m/s. } }

\item  {\footnotesize {FIG. 2 - X-ray spectra of $v-SiO_2$ at $Q$=1.6 nm$%
^{-1}$ taken at different temperatures. The data are shown together with
their best fits (full line) and the individual contributions to the fitting
function: elastic peak (dotted line), and inelastic components (dashed
line). In the inset are reported: 1) ($\bullet $) the inelastic-to-elastic
intensity ratio, $R(T)$; it follows the expected linear behaviour; and 2) ($%
\circ $) the excitation energy, $\Omega (T)$, derived from the DHO model.
The reported full lines are their linear fits. } }

\item  {\footnotesize {FIG. 3 - Temperature dependence of the scaling factor 
$a(T)$, normalized to $a(0)=1$, for the different spectra. Open symbols
refer to {\it incoherent} measurements of the BP: INS (open circles) and
Raman scattering (open squares). Full symbols are the scaling factors for
the excitations energies determined from the dynamics structure factor $%
S(Q,\omega )$ measured at $Q$=1.6 nm$^{-1}$ by IXS (full circles) and at $Q$%
=0.036 nm$^{-1}$ by BLS (full squares). The dashed lines are linear guides
to the eye to emphasise differences and similarities among the temperature
behaviours of the four data sets. } }

\item  {\footnotesize {FIG. 4 - Examples of RS spectra taken, from to top to
bottom, at 333, 523, 823, and 1073 K. The reported data correspond to the
reduced Raman intensities divided by the $C(\omega )$ (Fontana {\it et al}.
submitted) and shown in inset a). These spectra correspond to the $g(\omega
)/\omega ^2$. In inset b), the energy position of the maximum intensity at
each temperature ($\bullet $) is reported togheter with its linear fit in
the low temperature region (dashed line). } }
\end{description}

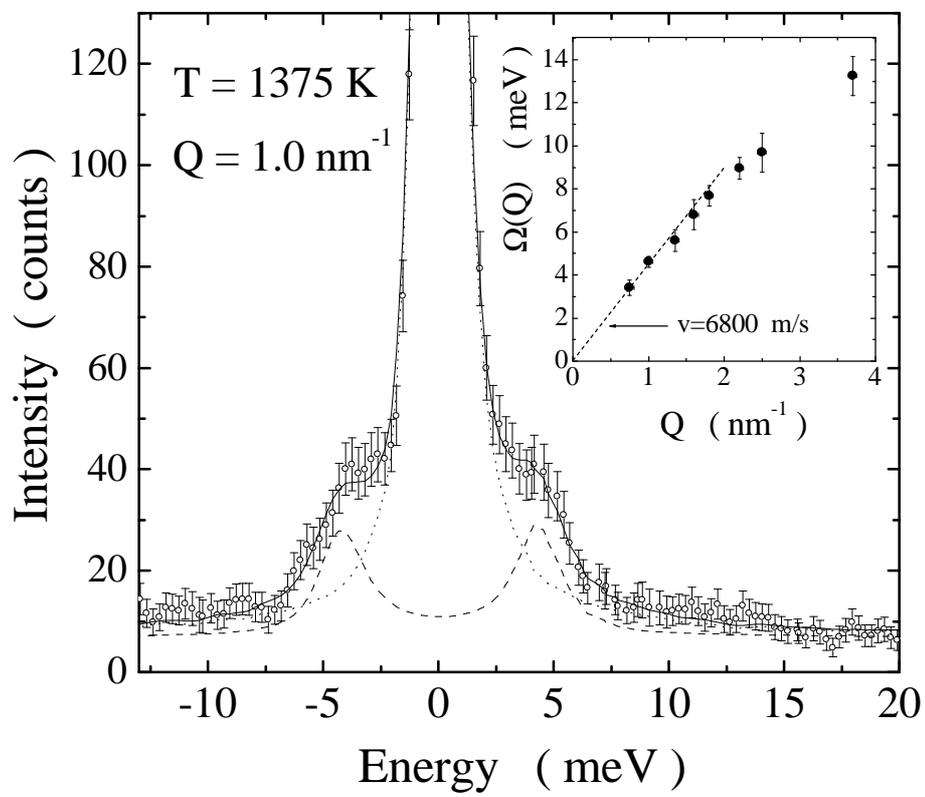
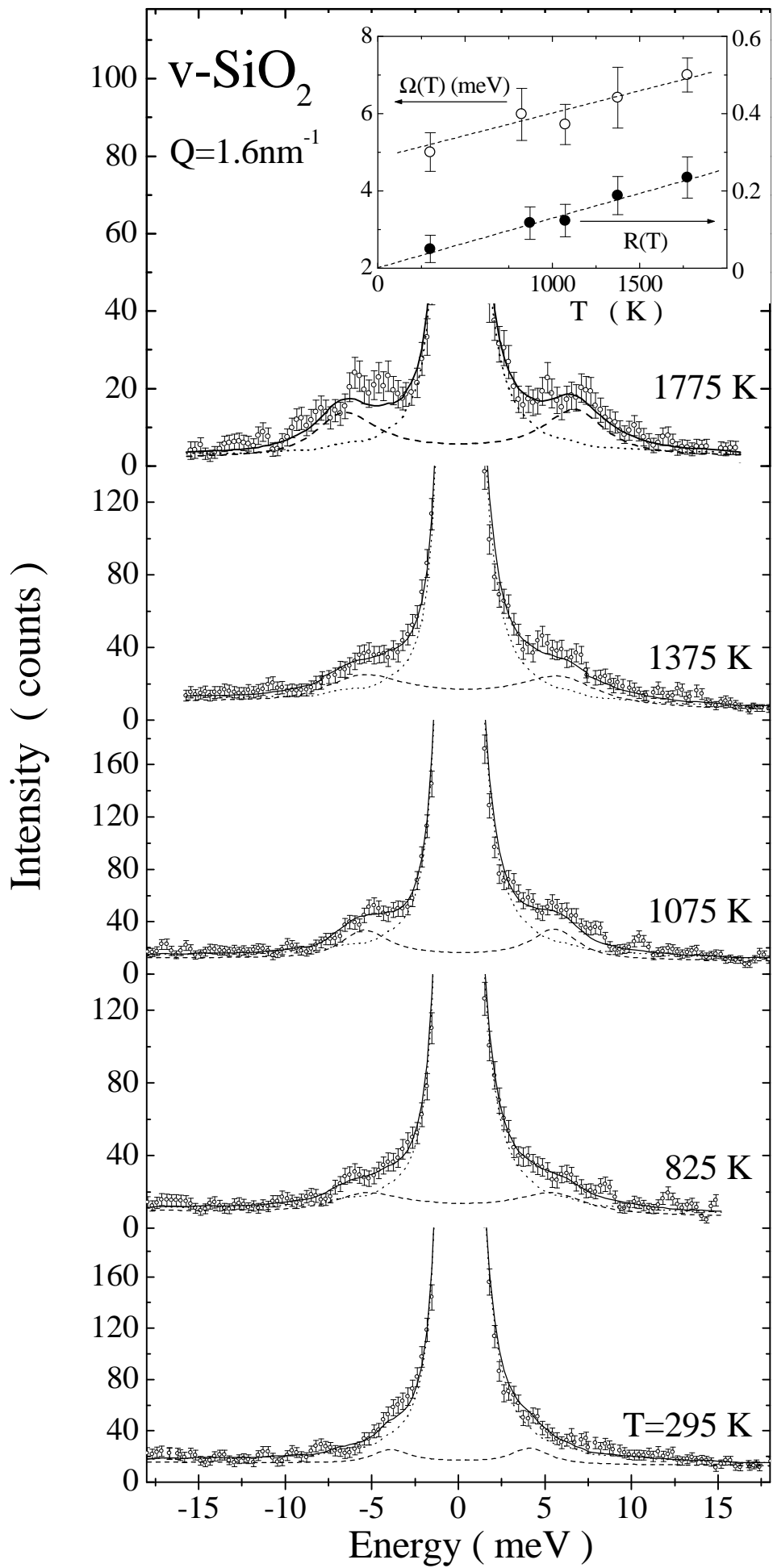
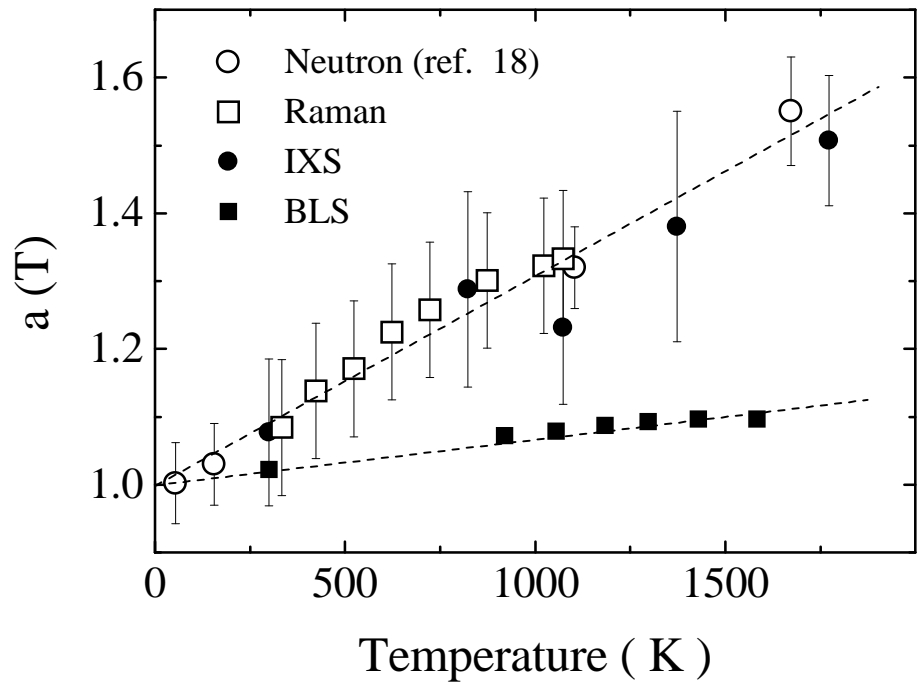
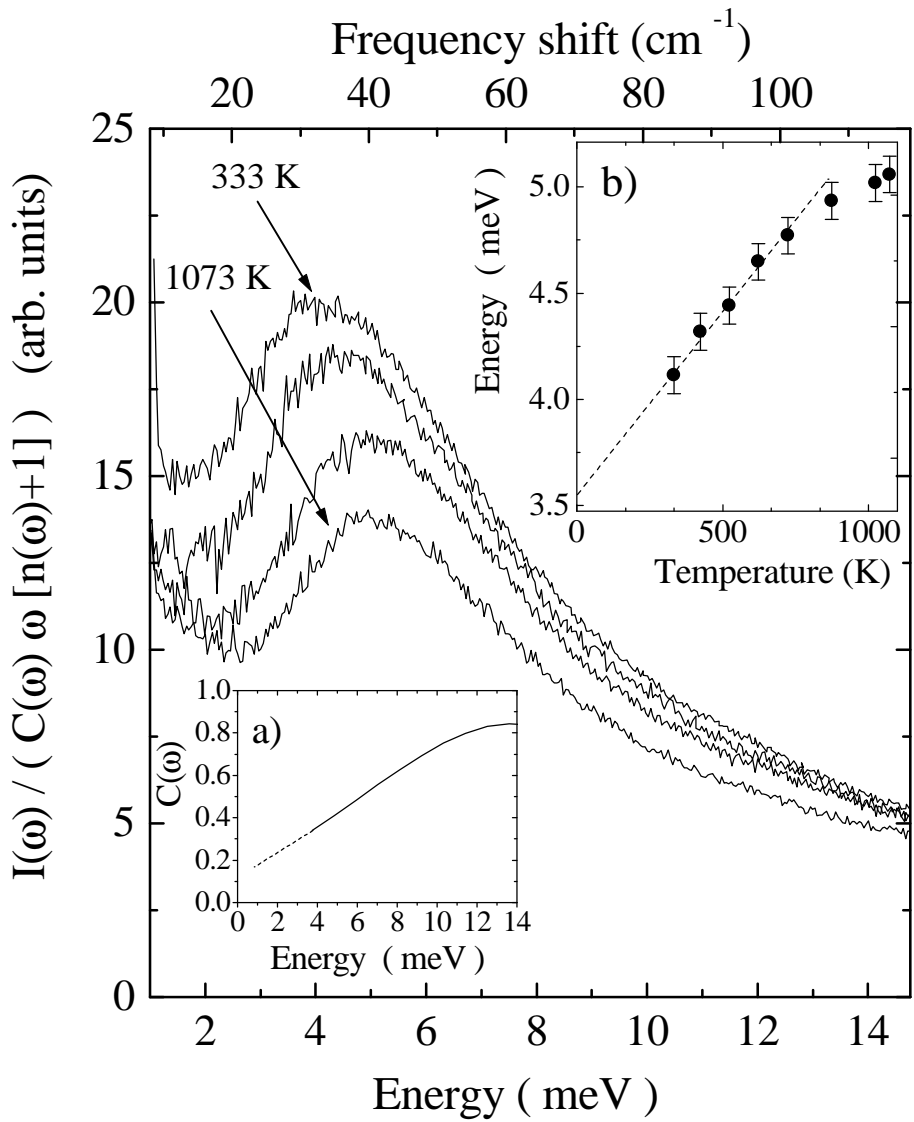
\end{document}